\documentclass{article}
\usepackage{spconf,amsmath,graphicx}
\usepackage{multirow}
\usepackage{stfloats}

\def\z{{\mathbf z}}

\def\T{{\mathbf T}}

\usepackage{enumitem}
\setenumerate[1]{itemsep=0pt,partopsep=0pt,parsep=\parskip,topsep=0pt}
\setitemize[1]{itemsep=0pt,partopsep=0pt,parsep=\parskip,topsep=0pt}
\setdescription{itemsep=0pt,partopsep=0pt,parsep=\parskip,topsep=0pt}

\title{Learning deep representations by multilayer bootstrap networks for speaker diarization}
\name{Meng-Zhen Li$^1$ and Xiao-Lei Zhang$^{1,2}$}
\address{$^1$CIAIC and School of Marine Science and Technology, Northwestern Polytechnical University, China \\
$^2$Research \& Development Institute of Northwestern Polytechnical University in Shenzhen, China \\
E-mail: lemonmengzhen@gmail.com, xiaolei.zhang@nwpu.edu.cn}
\begin{document}
\ninept
\maketitle
\begin{abstract}
  The performance of speaker diarization is strongly affected by its clustering algorithm at the test stage. However, it is known that clustering algorithms are sensitive to random noises and small variations, particularly when the clustering algorithms themselves suffer some weaknesses, such as bad local minima and prior assumptions. To deal with the problem, a compact representation of speech segments with small within-class variances and large between-class distances is usually needed. In this paper, we apply an unsupervised deep model, named multilayer bootstrap network (MBN), to further process the embedding vectors of speech segments for the above problem. MBN is an unsupervised deep model for nonlinear dimensionality reduction. Unlike traditional neural network based deep model, it is a stack of $k$-centroids clustering ensembles, each of which is trained simply by random resampling of data and one-nearest-neighbor optimization. We construct speaker diarization systems by combining MBN with either the i-vector frontend or x-vector frontend, and evaluated their effectiveness on a simulated NIST diarization dataset, the AMI meeting corpus, and NIST SRE 2000 CALLHOME database. Experimental results show that the proposed systems are better than or at least comparable to the systems that do not use MBN.
\end{abstract}
\section{Introduction}
\label{sec:intro}

Speaker diarization aims to solve the problem "who spoke when". It finds applications in speech recognition, communications, etc. A standard speaker diarization system contains four components---preprocessing, segmentation, frontend feature extraction, and clustering. 
Speech preprocessing removes silent segments by voice activity detection and sometimes reduces noise by speech enhancement \cite{sun2018novel}.
Speaker segmentation partitions an audio recording to a number of uniform short segments under a speaker-homogeneous assumption, or applies speaker change detection to partition the audio recording to nonuniform speaker-homogeneous segments \cite{hruz2017convolutional}. Frontend feature extraction applies speaker verification frontends to transform speaker segments into speaker embedding vectors, such as i-vectors \cite{dehak2010front}, d-vector \cite{garcia2017speaker,wang2018speaker,zajic2017speaker}, x-vector \cite{snyder2017deep,snyder2018x}, etc.
Speaker clustering partitions the speaker embedding vectors into several groups, each of which assumes to belong to a single speaker. Most research on speaker diarization focuses on the frontend feature extraction and speaker clustering components. Recently, an end-to-end speaker diarization that conducts the aforementioned four components by a neural network based system was also developed \cite{fujita2019end}.

This paper focuses on the speaker clustering component, since it not only affects the performance of speaker diarization significantly, but also is a key component that distinguishes speaker diarization from other speaker recognition tasks. We summarize some speaker clustering methods as follows. The most widely used clustering algorithm over the last decades was agglomerative hierarchical clustering (AHC) \cite{kenny2010diarization,maciejewski2018characterizing,sell2014speaker}. It works on the output feature space of frontends. Because AHC is sensitive to random noise, spectral clustering were frequently applied to reduce the irrelevant random noise of the speaker embedding vectors \cite{shum2012use,wang2018speaker,lin2019lstm}. It first projects the speaker embedding vectors, such as i-vectors \cite{shum2012use} or d-vectors \cite{wang2018speaker}, into a kernel-induced feature space, and then conducts Laplacian eigen-decomposition to reduce the random noise which transforms the speaker embedding vectors to new features for AHC. The kernel functions of spectral clustering are usually predefined, such as the Gaussian kernel, which limits its effectiveness. To make the spectral clustering more flexible, some works replaced the Gaussian kernel matrix to learnable affinity matrices, such as the affinity matrices produced from probability linear discriminant analysis (PLDA) or bidirectional long short-term memory networks \cite{lin2019lstm}. Besides AHC and spectral clustering, other common clustering methods were tried as well, such as the iterative mean-shift algorithm \cite{senoussaoui2014study}.

It is easy to see from the above analysis that reducing random noise is a core problem of speaker clustering. Here we propose an unsupervised deep learning method, named multilayer bootstrap network (MBN) \cite{zhang2018multilayer}, to tackle the problem.
MBN reduces the small variations and noise components of speaker embedding vectors steadily from bottom-up by building a gradually narrowed deep ensemble network. Each layer of MBN is a nonparametric density estimator based on random resampling of data and one-nearest-neighbor optimization. We further propose two MBN based speaker diarization systems that adopt the i-vector and x-vector frontends respectively.
 We have conducted a systematic comparison with the i-vector and x-vector speaker diarization systems that do not use MBN on both a synthetic NIST data and the real-world AMI meeting corpus and NIST SRE 2000 CALLHOME database. Evaluation results show that the proposed systems yield more compact representations than the conventional i-vectors and x-vectors; they also produce lower diarization error rates (DERs) than the comparison methods when the i-vector front-end is used.



\section{Proposed Method}
\label{sec:proposed method}

An overview of the proposed system is shown in Fig.\ref{fig:Overview}. After preprocessing audio data, such as silence removal and MFCC feature extraction, the proposed method first extracts speaker embedding vectors, including i-vectors and x-vectors, by a speaker recognition frontend. Then, it trains a PLDA model on the speaker embedding vectors. Finally, it applies MBN to transform the latent variables of PLDA to new speaker vectors, named m-vectors, and conducts clustering on the m-vectors by a conventional AHC. We present the details of the components as follows.

\subsection{Speaker embedding frontends}
\label{sssec:baseline}

\subsubsection{i-vector frontend}
\label{sssec:Representation}
The i-vector front-end contains a Gaussian mixture model based universal background model (GMM-UBM) $\Omega$ which is a speaker- and channel-independent GMM trained from the pool of all frame-level MFCC features of development data, and a total variability matrix $\T$ that encompasses both
speaker- and channel-variability. In the test stage, it first extracts a segment-level supervector from the frame-level features of each speech segment, and then reduces the dimension of the supervectors to i-vectors by the total variability matrix.

 \subsubsection{x-vector frontend}

The x-vector frontend is the state-of-the-art frontend for speaker verification. It is a time-delay neural network (TDNN) that consists of frame-level one-dimensional convolutional layers, a time-pooling layer, and segment-level fully connected layers from bottom-up. TDNN uses the time-pooling layer to transform frame-level features into segment-level speaker embeddings, and then uses the fully connected networks to classify the speaker embeddings. The three parts are jointly trained. In the test stage, the x-vectors are the output of the penultimate linear layer.

   \begin{figure}[t]
   \centering
   \includegraphics[width=0.4\textwidth]{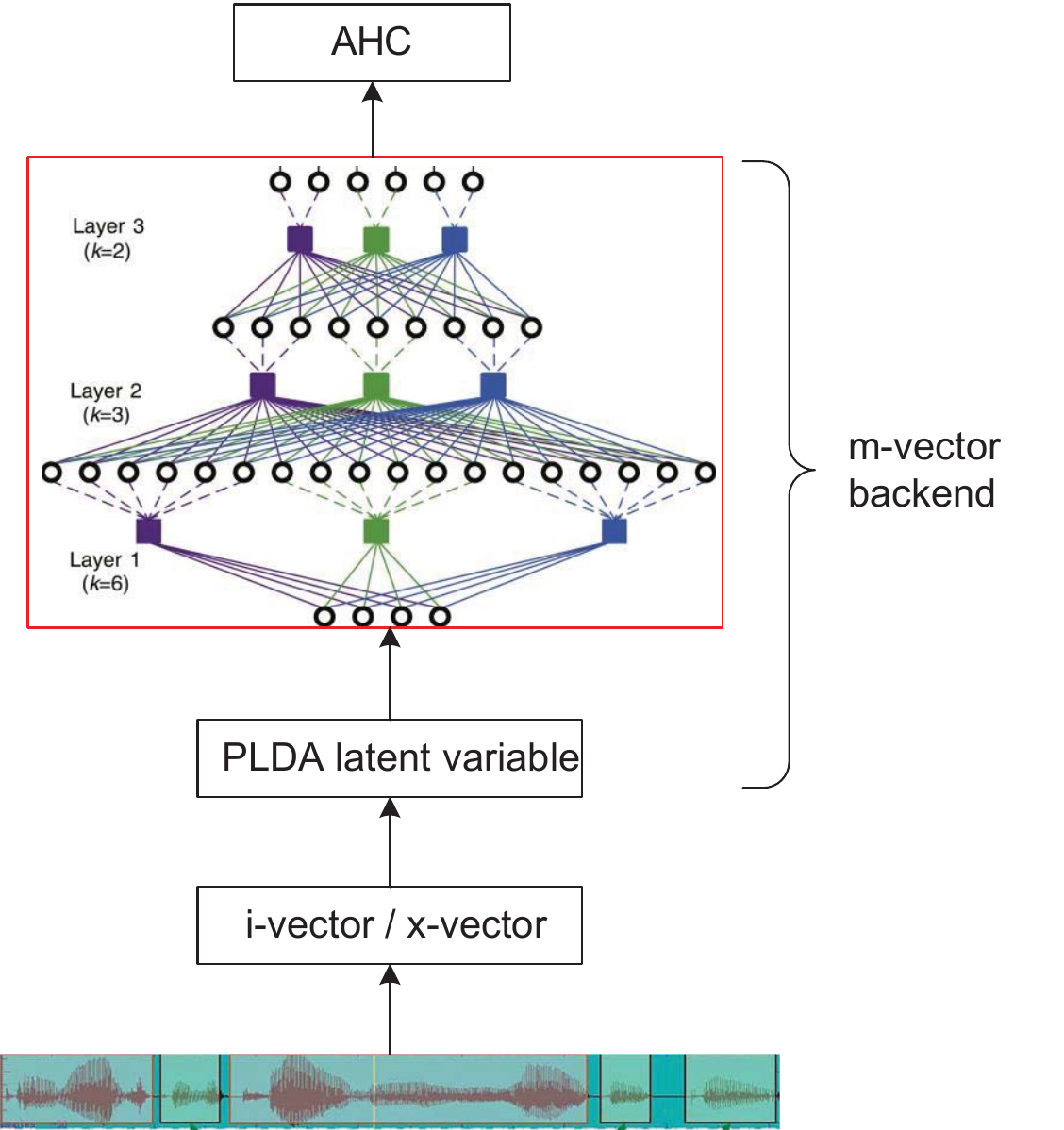}\\
   \caption{Diagram of the proposed system. The blue arrows represent baseline system procedures. The orange arrows represent proposed MBN based systems steps.  }
   \label{fig:Overview}
   \end{figure}

\subsection{m-vector backend}

\subsubsection{Latent variable extractor by PLDA}
\label{sssec:PLDA}

PLDA \cite{ioffe2006probabilistic} models the speaker embedding vector $\mathbf{x}$ by $P(\mathbf{x}|\mathbf{y})=\mathcal{N}(\mathbf{x}|\mathbf{y},\mathbf{\Phi}_w)$, where $\mathbf{y}$  is the class center which follows a Gaussian prior distribution $P(\mathbf{y})=\mathcal{N}(\mathbf{y}|\mathbf{m},\mathbf{\Phi}_b)$, with $ \mathbf{m} $ as the mean of the training data and $\mathbf{\Phi}_w$ and $\mathbf{\Phi}_b$ as the within-class and between-class covariances respectively. The above variables are estimated by the expectation maximization algorithm. According to \cite{ioffe2006probabilistic}, after the simultaneous diagonalization of the two covariance matrices, the PLDA model then becomes:
\begin{equation}\label{PLDA_mode}
   \mathbf{x} = \mathbf{m}+\mathbf{A}\mathbf{u}
\end{equation}
where $\mathbf{u}$ is the latent variable of $\mathbf{x}$, and $\mathbf{A}\mathbf{A}^T=\mathbf{\Phi}_w$. We take $\mathbf{u}$ as the input of MBN.

\subsubsection{m-vector extractor by MBN}

MBN is used in the test stage only. It transforms the latent variables of PLDA $\mathbf{u}$ into low-dimensional m-vectors.
As illustrated in the red box of Fig. \ref{fig:Overview}, MBN contains multiple hidden layers. Each hidden layer consists of $V$ independent \textit{k}-centroids clusterings, where $V\gg 1$. Each \textit{k}-centroids clustering has \textit{k} output units, each of which indicates a cluster. The output units of all \textit{k}-centroids clusterings in the same layer are concatenated as the input of their upper layer. Different from the original work in \cite{zhang2018multilayer}, we discard the PCA layer, and use the output of the last hidden layer as the output of MBN.

MBN is a gradually narrowed network built layer-by-layer from bottom-up. Suppose MBN contains $L$ layers, and the parameters $k$ from the bottom hidden layer to the top hidden layer are denoted as $k_1,\ldots, k_L$ respectively. The parameters $k_1,\ldots, k_L$ are determined by the following criteria:
\begin{eqnarray}
 &k_1 \gg O, \label{eq:k1}\\
 & k_{l+1}=\delta k_l,\label{eq:rel}\quad \forall l = 1,\ldots,L-1,\\
 & k_L \mbox{is set to eusure at least one data point per class in probability}\label{eq:kL}
\end{eqnarray}
where $k_1$ and $\delta\in [0,1)$ are user-defined hyperparameters, and $O$ is the ground-truth number of speakers in the test utterance. It can be seen that $L$ is determined automatically. Note that the criterion \eqref{eq:kL} is usually specified to $k_L \geq  \left\lceil 1.5 O\right\rceil$ for class-balanced problems, and simply set to a large number, e.g. 100, for severely class-imbalanced problems.


 For training each layer given a $d$-dimensional input data set $\mathcal{Z} = \left\{\mathbf{z}_1,\ldots,\mathbf{z}_n \right\}$ either from the lower layer or from PLDA, MBN trains each $k$-centroids clustering independently via the following two steps \cite{zhang2018multilayer}: (i) The first step randomly selects $k$ data points from $\hat{{\mathcal{Z}}}$ as the $k$ centroids of the clustering, denoted as $\{\mathbf{w}_{1},\ldots,\mathbf{w}_{k} \}$. (ii) The second step learns a sparse vector $\mathbf{h}$ from the input $\z$. $\mathbf{h}$ is a one-hot code indicating the nearest centroid of ${\z}$. For example, if the third centroid is the nearest one to ${\mathbf{z}}$, then $\mathbf{h} = [0,0,1,0,\ldots,0]^T$. The similarity metric between the centroids and ${\mathbf{z}}$ at the bottom layer is the log likelihood ratio adopted by the PLDA scoring. The similarity metric at all other layers are set to $\arg\max_{i=1}^{k}\mathbf{w}_i^T{\mathbf{z}}$.

The m-vector produced from the top layer of MBN is:
\begin{eqnarray}
\mathbf{m} = [\mathbf{h}_1^T,\mathbf{h}_2^T,\ldots, \mathbf{h}_V^T]^T
\end{eqnarray}
where $\mathbf{h}_v$ is the sparse output of the $v$th $k_L$-centroids clustering.  

\subsubsection{Speaker clustering by AHC}

We compute the similarity scores between the m-vectors $\{\mathbf{m}_i\}_{i=1}^n$ by the cosine similarity:
   \begin{equation}
d_{\rm{cos}}(\mathbf{m}_i, \mathbf{m}_j) = \frac{\mathbf{m}_i^T\mathbf{m}_j}{\|\mathbf{m}_i\|_2 \|\mathbf{m}_j\|_2},\quad \forall i,j = 1,\ldots,n
\end{equation}
and use the average linkage AHC to partition the m-vectors into groups for speaker diarization.



\section{Experiment}
\label{sec:experiment}

\subsection{Data sets}

 We first conducted an experiment on a controllable simulated data, and then evaluated the proposed systems on the real-world AMI meeting corpus and CALLHOME dataset.

\underline{\textbf{Synthetic test:}}
We used the 8 conversation female data of NIST SRE 2006 as the training data, and the 8 conversation female data of NIST SRE 2008 as the test data, which consist of 402 and 395 female speakers respectively. Each speaker contains 8 utterances. Every utterance is about 2 minutes long after removing the silence regions. We merged the speech of every five speakers in the test data into a conversation, which amounts to 79 conversations. We can see that this is an ideal test environment without the problems of overlapping or over short speech segments and over short conversations.

\underline{\textbf{Real-world test:}}
The training data is the switchboard data and a collection of NIST SREs, including NIST SRE 2004, 2005, 2006, 2008. We used the training data to train the i-vector frontend and PLDA model. We further augmented the training data with reverberation, noise, music, and babble that are generated from room impulse response functions and the MUSAN corpus, and combined the augmented data with the clean data to train the x-vector frontend and PLDA model.

The first evaluation dataset is the AMI meeting corpus which contains about 100 hour meeting recordings. It uses a single microphone from a tabletop array to form a synthetic far-field diarization condition. Each meeting recording consists of four to five speakers, and lasts about thirty minutes.
The corpus consists of a development set and an evaluation set, each of which amounts to about 10$\%$ of the full corpus. We merged the two sets as our test set.
The remaining 80$\%$ of the corpus is reserved for training the i-vector frontend.
 

The second evaluation dataset is the CALLHOME dataset. Each recording of the CALLHOME dataset is a single-channel telephone conversation, which contains two to seven speakers. The corpus contains six languages: Arabic, English, German, Japanese, Mandarin and Spanish. It consists of 500 short conversations, each of which lasts about 1 to 10 minutes. The average time of the conversations is about two minutes.

\subsection{Experimental setup}
\label{sssec:Experimentalsetup}

We used a sliding window with a window size of 1.5 seconds and a window shift of 0.75 second to partition speech recordings into speech segments, each of which is 1.5 seconds long. We extracted 20-dimensional MFCC for the i-vector frontend and 23-dimensional MFCC for the x-vector frontend. The i-vector frontend contains a GMM-UBM of 2048 Gaussian components, and a total variance matrix of 128 dimensions. The x-vector frontend in use is the same as the x-vector implementation in Kaldi, which outputs 128-dimensional x-vectors \cite{sell2018diarization}. 
We took the Kaldi callhome diarizaton system\footnote{https://github.com/kaldi-asr/kaldi/tree/master/egs/callhome\_diarization} as the baseline system. The baseline system and the proposed method used the same frontends.

We report the hierarchical clustering results under two circumstances: \textit{oracle} and \textit{thresholding}. The oracle circumstance assumes that the speaker number of a conversation is known as a prior. The thresholding circumstance does not have this prior. To determine the number of speakers on-the-fly, we followed the Kaldi's implementation which uses a development set to determine the threshold of AHC.
Eventually, the threshold was chosen from a range of $[-0.3,0.3]$ for each conversation.

We evaluated the effectiveness of the proposed method from the following three aspects. The first aspect evaluates the discriminant ability of the m-vectors by visualizing them in a two dimensional subspace of PCA.
%
The second aspect is to compute the following discriminant trace (DT) criterion \cite[eq. (4.50)]{bishop2006pattern} ${\rm{DT}}=\operatorname{Tr}\left\{\mathbf{s}_{\mathrm{B}}^{-1} \mathbf{s}_{\mathrm{W}}\right\}$,
where $ \mathbf{s}_{\mathrm{B}} $ is the between-class covariance and $ \mathbf{s}_{\mathrm{W}} $ is the within-class covariance. When $ \mathbf{s}_{\mathrm{B}} $ is large and when $ \mathbf{s}_{\mathrm{W}} $ is small, the ${\rm{DS}}$ score is small, which means that the feature representation is good. 
The third aspect is to evaluate the diarization result in terms of DER. DER was described and used by NIST in the rich transcription evaluations (NIST Fall Rich Transcription on meetings 2006 Evaluation Plan, 2006). It measures the fraction of time that is not classified correctly to a speaker or to non-speech. We skip the overlap speech. Eventually, DER is calculated by ${\rm{DER}} = E_{\rm{spkr}} + E_{\rm{MISS}} + E_{\rm{FA}}$ where  $E_{\rm{spkr}}$, $E_{\rm{MISS}}$, and $E_{\rm{FA}}$ are the speaker error rate, miss-detected speech, and false alarm speech respectively.


   \begin{figure}[t]
   \centering
   \includegraphics[width=0.42\textwidth]{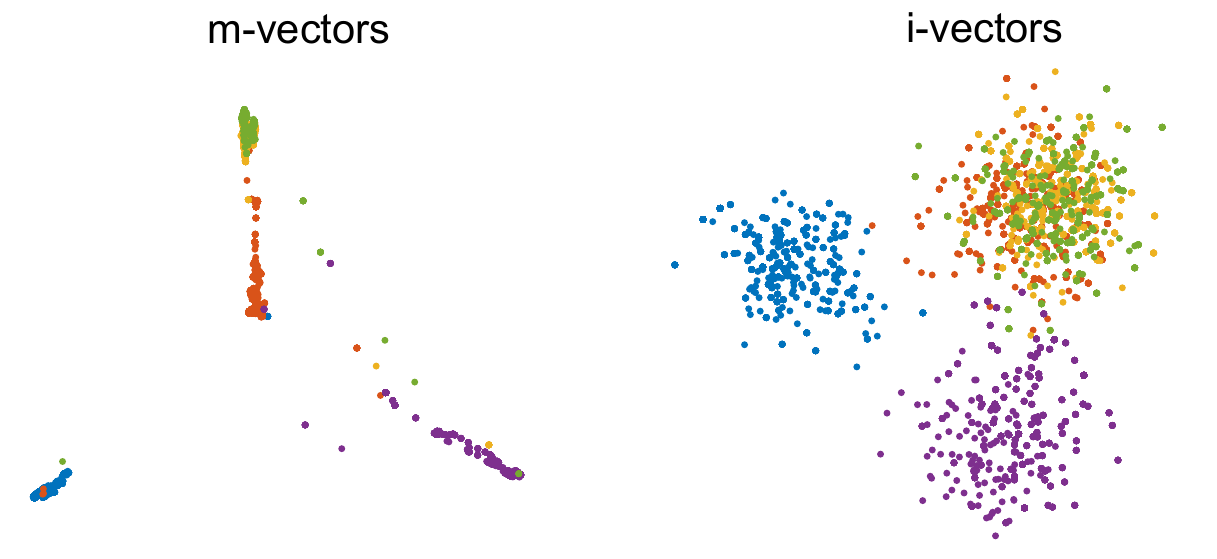}\\
   \caption{Visualizations of m-vectors and i-vectors produced by PCA on the synthetic data, when the i-vector frontend is used. Each data point represents a speaker segment of 1.5 seconds long. Different colors represent different speakers.}
   \label{fig:visual1}
   \end{figure}

   \begin{figure}[t]
   \centering
   \includegraphics[width=0.42\textwidth]{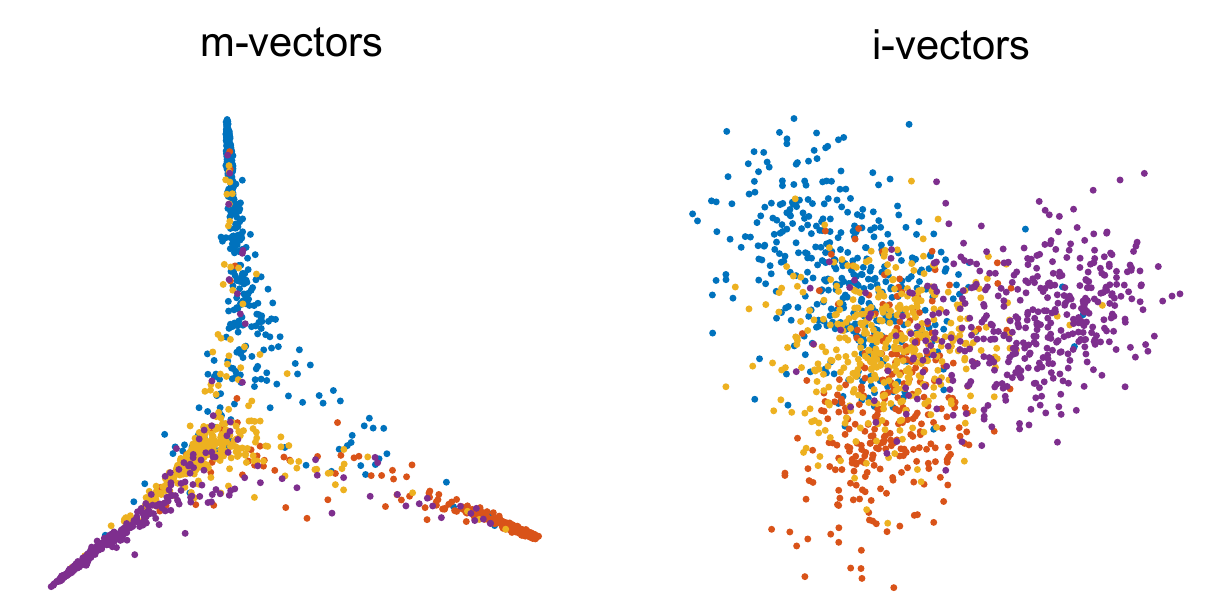}\\
   \caption{Visualizations of m-vectors and i-vectors on the AMI corpus, when the i-vector frontend is used. For simplifying the visualization demo, the speaker segments containing multiple speakers were assigned to the first speaker.}
   \label{fig:visual2}
   \end{figure}

      \begin{table}[t]
\centering
\caption{DT scores of the i-vector baseline and the proposed system with the i-vector frontend on the synthetic data and AMI corpus.}
\label{tab:xxx1}
\scalebox{1}{
\begin{tabular}{llrr}
\hline
 Data & Method &  \\ \hline
 \multirow{2}{*}{Synthetic} & baseline & 4.11 \\
 & proposed & \textbf{0.22}  \\ \hline
\multirow{2}{*}{AMI} & baseline & 1.68 \\
 & proposed & \textbf{1.34}  \\ \hline
\end{tabular}}
\end{table}

   \begin{table}[t]
\centering
\caption{DER results of the i-vector baseline and the proposed system with the i-vector frontend on the three evaluation data sets.}
\label{tab:der1}
\scalebox{1}{
\begin{tabular}{llrr}
 \hline
Data & Method & Oracle & Thresholding \\ \hline
 \multirow{2}{*}{Synthetic} & baseline & 23.60$\%$ &--- \\
 & proposed & \textbf{7.42\%} & --- \\ \hline
\multirow{2}{*}{AMI} & baseline & 29.20$\%$ & 31.88$\%$ \\
 & proposed & \textbf{23.38\%} & \textbf{24.08\%} \\ \hline
\multirow{2}{*}{CALLHOME} & baseline &  \textbf{8.92\%} & \textbf{10.71\%} \\
 & proposed & \textbf{9.07\%} & 11.41$\%$ \\ \hline
\end{tabular}}
\end{table}
%


\subsection{Results with the i-vector frontend}
\label{sssec:results}

Both of the proposed method and the baseline system in this subsection use the i-vector front-end. The experiments were conducted on both the synthetic data and two real-world datasets. The hyperparameters of MBN were set to $ V = 400 $, $k_1 = 50$, and $\delta=0.3$ for the synthetic data and AMI corpus, and were set to $ V = 400 $, $k_1 = 10$, and $\delta=0.3$ for the CALLHOME database. The main reason why $k_1$ for the CALLHOME data was set to a small value is that CALLHOME contains only short conversations. Setting a large $k_1$ not only wastes computational power but also has a risk of overfitting to noise. We present the experimental details as follows.

We first conducted experiments on the synthetic data and AMI corpus. 
Figures \ref{fig:visual1} and \ref{fig:visual2} show the visualizations of the m-vectors and i-vectors on the two corpora respectively. From the two figures, we see clearly that the distributions of the m-vectors have smaller overlaps than the distributions of the i-vectors. Table \ref{tab:xxx1} lists the DT scores of the m-vectors and i-vectors. From the table, we see that the m-vector representation has a strong discriminability than the i-vector representation. Table \ref{tab:der1} lists the DER results produced by the comparison systems. From the table, we see that the proposed system achieves an over 15\% absolute DER reduction over the i-vector baseline on the synthetic data, and about 7\% absolute DER reduction over the latter on the AMI corpus in both the oracle and thresholding circumstances. Because MBN is relatively sensitive to the hyperparameter $ \delta $ \cite{zhang2018multilayer}, we studied $\delta$ on the synthetic data by a grid search from $\{0.3, 0.5, 0.7\}$. The DER scores are 7.42\%, 7.19\%, and 8.09\% respectively, which means that different $ \delta $ only slightly affects the performance. In practice, we usually set $\delta$ to a small value, such as 0.3, since that, when $\delta$ is set to a small value, MBN is used mainly for denoising; otherwise, MBN is used for the reduction of nonlinearity.

In addition, we evaluated the diarization performance of the comparison methods on the CALLHOME dataset.
From the result in Table \ref{tab:der1}, we see that the DER scores of the proposed system were slightly lower than those of the baseline system. The reason why the proposed system does not outperform the baseline system might be that the conversations are too short to demonstrate the advantage of deep and nonparametric methods. The average time of the conversations can only be split into no more than 200 speech segments. Note that, because we do not have the ground-truth labels of the speech segments, we are unable to obtain the DT scores of the comparison methods on CALLHOME.


\begin{figure}[t]
   \centering
   \includegraphics[width=0.42\textwidth]{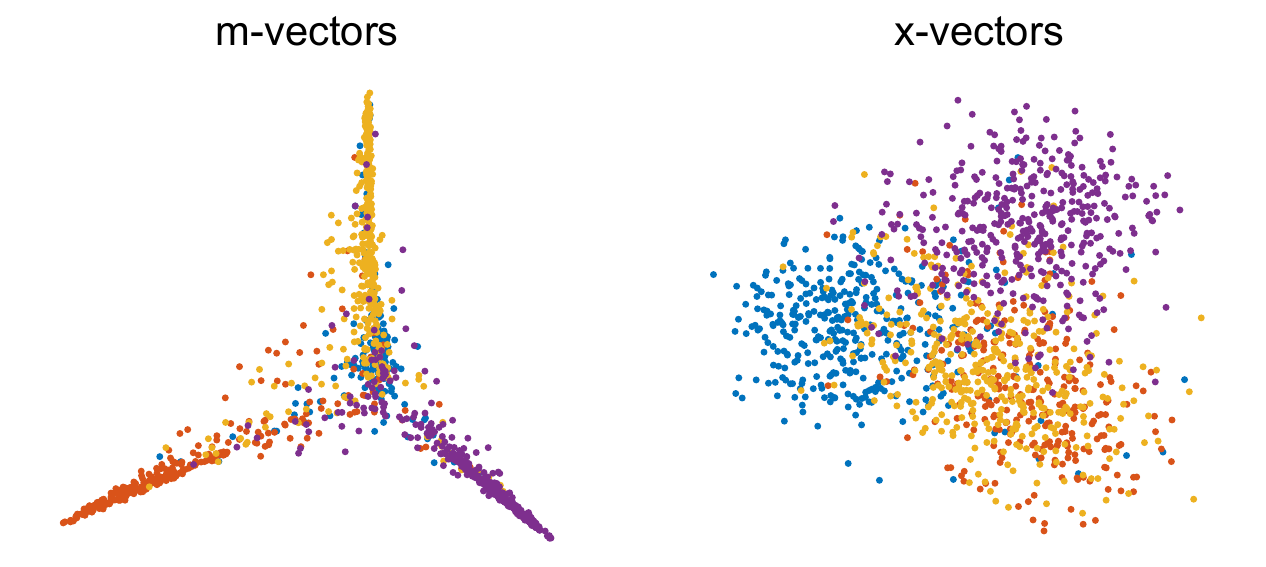}\\
   \caption{Visualizations of m-vectors and x-vectors on the AMI corpus, when the x-vector frontend is used.}
   \label{fig:visual3}
   \end{figure}

  \begin{table}[t]
\centering
\caption{DT scores of the x-vector baseline and the proposed system with the x-vector frontend on the AMI corpus.}
\label{tab:xxx2}
\scalebox{1}{
\begin{tabular}{lllrr}
\hline
 Data &Method &  \\ \hline
\multirow{2}{*}{AMI} & baseline & \textbf{4.36} \\
 & proposed & 6.50  \\ \hline
\end{tabular}}
\end{table}

  \begin{table}[hb]
\centering
\caption{DER results of the x-vector baseline and the proposed system with the x-vector frontend on AMI and CALLHOME.}
\label{tab:der2}
\scalebox{1}{
\begin{tabular}{llrr}
 \hline
Data &Method & Oracle & Thresholding \\ \hline
\multirow{2}{*}{AMI} & baseline &  \textbf{29.76\%} & \textbf{30.10\%} \\
 & proposed & \textbf{29.96\%} & 30.64\% \\ \hline
\multirow{2}{*}{CALLHOME} & baseline &  7.15$\%$ & \textbf{8.73\%} \\
 & proposed & \textbf{6.22\%} & \textbf{8.64\%}  \\ \hline
\end{tabular}}
\end{table}

\subsection{Results with the x-vector frontend}

We evaluated the proposed method with the x-vector frontend on the AMI corpus, where we set the hyperparameters of MBN the same as those with the i-vector frontend. Figure \ref{fig:visual3} shows the visualization results of the m-vectors and x-vectors. From the figure, we see that the proposed m-vectors seem have smaller within-class variance and between-class distances than the x-vectors. However, comparing Fig. \ref{fig:visual2} with Fig. \ref{fig:visual3}, we see that the the m-vectors with the x-vector frontend are not as good as those with the i-vector frontend. The results in Tables \ref{tab:xxx2} and \ref{tab:der2} further manifest the above experimental phenomena. The proposed system is not as good as the x-vector baseline in terms of DT, and performs equivalently well with the latter in terms of DER. These phenomena might be caused by the mismatch between the training and test. Unlike the i-vector frontend where we have used 80\% AMI data as part of the training data, we did not use any AMI data for training the x-vector frontend. However, comparing Table \ref{tab:der1} with Table \ref{tab:der2}, we see that the proposed system with the i-vector frontend achieves the best performance.

We further evaluated the proposed method on the CALLHOME corpus. The experimental phenomenon in Table \ref{tab:der2} is similar with that in Table \ref{tab:der1}.

%
%

%
%

\section{Conclusions}
\label{sec:Conclusions}

In this paper, we apply multilayer bootstrap networks to reduce the noise and small variance of speaker embedding vectors. Specifically, it first extracts speaker embedding vectors, such as i-vectors or x-vectors, from a front-end. Then, it transforms speaker embedding vectors to latent variables of PLDA. Finally, it uses the latent variables as the input of MBN, which produces high-dimensional sparse m-vectors for AHC. A key advantage of MBN is that it builds an unsupervised deep model without model assumptions and difficult optimization algorithms. We have conducted a wide experimental comparison with the i-vector and x-vector based speaker diarization systems on a simulated NIST diarization dataset, the AMI Meeting Corpus, and NIST SRE 2000 CALLHOME database. Experimental results show that our systems are better than or at least comparable to the systems that do not adopt MBN.

%



\bibliographystyle{IEEEbib}
\bibliography{strings,refs}

\end{document}